# COMPUTATIONAL ANALYSIS OF FACTORS INFLUENCING ENHANCEMENT OF THERMAL CONDUCTIVITY OF NANOFLUIDS

George Okeke, Sanjeeva Witharana, Joseph Antony*, Yulong Ding

Institute of Particle Science and Engineering, University of Leeds, Leeds, LS2 9JT, UK

*S.J.Antony@leeds.ac.uk

**ABSTRACT**

Numerical investigations are conducted to study the effect of factors such as particle clustering and interfacial layer thickness on thermal conductivity of nanofluids. Based on this, parameters including Kapitza radius, and fractal and chemical dimension which have received little attention by previous research are rigorously investigated. The degree of thermal enhancement is analysed for increasing aggregate size, particle concentration, interfacial thermal resistance, and fractal and chemical dimensions. This analysis is conducted for water-based nanofluids of Alumina ($Al_2O_3$), CuO and Titania ($TiO_2$) nanoparticles where the particle concentrations are varied up to 4vol%. Results from the numerical work are validated using available experimental data. For the case of aggregate size, particle concentration and interfacial thermal resistance; the aspect ratio (ratio of radius of gyration of aggregate to radius of primary particle, $Rg/a$) is varied between 2 to 60. It was found that the enhancement decreases with interfacial layer thickness. Also the rate of decrease is more significant after a given aggregate size. For a given interfacial resistance, the enhancement is mostly sensitive to $Rg/a$ <20 indicated by the steep gradients of data plots. Predicted and experimental data for thermal conductivity enhancement are in good agreement.

On the influence of fractal and chemical dimensions ($d_l$ and $d_f$) of Alumina-water nanofluid, the $Rg/a$ was varied between 2-8, $d_l$ between 1.2-1.8 and $d_f$ between 1.75-2.5. For a given concentration, the enhancement increased with the reduction of $d_l$ or $d_f$. It appears a distinctive sensitivity of the enhancement to $d_f$, in particular in the range 2-2.25, for all values of $Rg/a$. However the sensitivity of $d_l$ was largely depended on the value of $Rg/a$. The information gathered from present work on the sensitivity of thermal conductivity enhancement to aggregate size, particle concentration, interfacial resistance, and fractal and chemical dimensions will be useful in manufacturing highly thermally conductive nanofluids. Further research on the refine cluster evolution dynamics as a function of particle-scale properties is underway.

*Keywords: thermal conductivity, nanofluids, aggregation, fractal dimensions, Titania, Alumina*

# INTRODUCTION

The concept of nanofluids refers to making stable suspensions of nanometer-size particles in conventional liquids. Particles could be made of metals, metal oxides or ceramics which are usually smaller than 100nm in size. Commonly used nanomaterials for making nanofluids are Aluminium oxide, Titanium dioxide, Silicon oxide and Carbon nanotubes. Popular host liquids are water, ethylene glycol, polyalfa olefin and propylene glycol. One of the advantages of nanofluids is their enhanced ability to conduct heat in heat transfer applications. It has been experimentally shown in many instances, that the nanofluids have displayed higher thermal conductivity and convective heat transfer in comparison to host liquids (Chen et al., 2008; Ding et al., 2010; Garg et al., 2009; Li and Peterson, 2006; Wen et al., 2006). However when the particle concentration is very low, the enhancement is negligible or very small (Kim et al., 2007; Putnam et al., 2006; Witharana et al., 2010). The latter observation was confirmed by the recently concluded International benchmarking exercise for nanofluids (J. Buongiorno et al., 2009). With regard to forced convective heat transfer, nanofluids had displayed more consistent data. Both in laminar and turbulent regimes, they appear to have enhanced the heat transfer (He et al., 2007; Wen and Ding, 2004; Xuan and Li, 2000).

The experimentally observed heat transfer enhancement caused by suspended nanoparticles has prompted deeper investigations into the subject. It was hypothesized that proper understanding of the underlying mechanisms would pave the way to custom-built nanofluids which will in-turn revolutionize energy conservation efforts. Hence the experiments were followed by numerical modeling and vice-versa. Soon it was reported that classical effective medium theories such as Maxwell (1881), Bruggeman (1935) and Hamilton-Crosser (1962) were unable to predict the experimental data very well. As a result, a number of hypotheses were put forward to explain the newly found behaviour of nanofluids.

The hypothesis of Keblinski et al (2002) laid foundation to most of the thermal conductivity prediction models published so far. They hypothesized that the observed data could be explained by the combination of four mechanisms. These are; Brownian motion of nanoparticles in the liquid, formation of liquid layers around nanoparticles, ballistic nature of heat transport within nanoparticles, and formation of nanoparticle clusters. This statement was followed by publication of several numerical models. Each model addressed one or more of the four mechanisms. Among them, there were widely discussed models such as Jang and Choi (2004), Ko and Kleinstreuer (2004), Kumar et al (2004) and Prasher et al (2005, 2006a; 2006b; 2006c). These models predict sections of thermal conductivity data but none of them is able to fully capture majority of data published in literature, although among the above said mechanisms, the formation of nanoparticle clusters is believed to be the most thermally influential mechanism. Further investigation including a new model that reviews existing models and addresses the deficiencies in them is required.

In the present work, numerical investigations are conducted in detail to study the effect of particle clustering and interfacial layer thickness on thermal conductivity of nanofluids. The interplay between these two factors is examined through Kapitza radius (Every *et al.*, 1992). The degree of enhancement is analysed for increasing aggregate size and particle concentration. The numerical work is validated with experimental data on water-based nanofluids of metals and metal oxide nanoparticles. On top of this, studies are conducted on the effect of fractal and chemical dimensions of nanoparticle clusters. The concept of fractal dimension has been used in the past to study the aggregation process of particles. Using this concept it has been ascertained that aggregate structures can be characterised by fractal and chemical dimension of objects as these structures are usually complex (2003). In simple terms, fractal dimension can be taken as the measure of structures or fractals in their decomposed self-similar state. It reveals the contributive and complex nature of self-similar pieces as they fill up space to form a single structure. In the same manner, chemical dimension can be taken as the minimum cluster path between two points (Margolina, 1985). A previous study by Waite et al (2001) on aggregated alumina nanoparticle suspension have found fractal dimension to range from 1.8 to 2.3. Prasher et al (2006c) indicates that fractal dimension of 2.5 signifies reaction-limited aggregation and is a strong repulsive barrier while that of 1.8 signifies diffusion-limited cluster-cluster aggregation (DLCCA) and is a weak repulsive barrier. From Wang et al (2003), aggregation in nanofluids have fractal dimensions close to 1.8 and therefore is DLCCA. Following these findings, calculations in other literatures such as Prasher et al (2006c) and Evans et al (2008) are based on fractal dimension of 1.8. The same applies to chemical dimension in which a value of 1.4 is fixed for calculations conducted by Evans et al (2008). The sensitivity of fractal and chemical dimension on the degree of enhancement is analyzed for varied aggregate sizes and particle concentration.

**COMPUTATIONAL ANALYSIS**

Aggregates are consisting of particles some of which are arranged like chains (also known as backbones) while the others remain as individual particles (also known as dead ends) (Evans *et al.*, 2008). For modeling purposes an aggregate is embedded in an imaginary sphere of radius $R_g$. Following the work of Evans et al (2008), numerical investigations are conducted hereby using the basic set of equations shown below.

$$k_{eff}/k_f = ([k_a + 2k_f] + 2\phi_a[k_a - k_f]) / ([k_a + 2k_f] - 2\phi_a[k_a - k_f]) \qquad (1)$$

Where $k_f$, $k_a$, $\phi_p$ and $\phi_a$ are thermal conductivities of the base fluid and particle aggregate, and, particle volume fraction and aggregate volume fraction respectively. The term $k_a$ is determined following Nan et al(1997)'s model for randomly oriented particles.

$$k_a = k_{nc} \frac{3 + \phi_c[2\beta_{11}(1-L_{11}) + \beta_{33}(1-L_{33})]}{3 - \phi_c[2\beta_{11}L_{11} + \beta_{33}L_{33}]} \qquad (2)$$

Where $k_{nc}$ is the thermal conductivity of the imaginary medium in which the backbones are embedded in. $\varphi_c$ is the volume fraction of backbone particles. Other parameters are defined in equations 3 to 6.

$$L_{11} = 0.5p^2/([p^2-1]) - 0.5p\ \cosh^{-1} p/[p^2-1] \qquad (3)$$

$$L_{33} = 1 - 2 L_{11} \qquad (4)$$

$$\beta_{ii} = (k_{ii}^c - k_{nc}) / [k_{nc} + L_{ii}(k_{ii}^c - k_{nc})] \qquad (5)$$

$$k_{ii}^c = k_p / (1 + \gamma L_{ii} k_p / k_f) \qquad (6)$$

Where $p$ is the aspect ratio for the aggregate defined by $Rg/a$. Interfacial thermal resistance is accounted for by a non-dimensional parameter $\alpha$ in a relation $\gamma = (2 + 1/p)\ \alpha$. $\alpha$ which relates interfacial thermal resistance to particle radius $a$, is given by $\alpha = a_k/a$, where $a_k$ is Kapitza radius which Every et al.(1992) defined in terms of the interfacial thermal resistance ($R_{Bd}$) and fluid thermal conductivity ($k_f$). It follows that $a_k = R_{Bd} \cdot k_f$. The chemical and fractal dimensions ($d_l$ and $d_f$ respectively) are accounted for using; $\phi_p = \phi_{int}\phi_p$, where $\phi_{int}$ is the volume fraction of the nanoparticles in a single aggregate, $\phi_{int} = (Rg/a)^{df-3}$. The chemical dimension, $d_l$, defines the number of particles belonging to the backbone, $N_c = (Rg/a)^{dl}$ where $d_l$ ranges between one and $d_f$. In the situation where all the particles belong to the backbone without any dead ends contained, it becomes $d_l = d_f$ and the volume fraction of the backbone particles $\phi_c$ in the aggregate is $\phi_c = (Rg/a)^{dl-3}$. $\phi_{nc}$ is taken as the volume fraction of the particles which belong to dead ends and is given as, $\phi_{nc} = \phi_{int} - \phi_c$. What remains is to calculate the term $k_{nc}$. From Evans et al (2008), it can be written as,

$$(1-\phi_{nc})(k_f - k_{nc}) / (k_f + 2k_{nc}) + \phi_{nc}(k_p - k_{nc})/(k_f + 2k_{nc}) = 0 \qquad (7)$$

Uniqueness of this model is the incorporation of all these parameters to be investigated into the model for the purpose of a more rigorous and comprehensive analysis.

Based on the model given above, numerical investigations were conducted to study the effect of particle clustering, interfacial layer thickness, particle concentration, fractal and chemical dimensions (respectively represented by $Rg/a$, $\alpha$ and $a_k$, $\varphi$, $d_f$, $d_l$) on the thermal conductivity of nanofluids. Finally the model predictions were compared with limited experimental data in literature for water based $Al_2O_3$, CuO and $TiO_2$ nanofluids.

**RESULTS**

To study the influence of interfacial thermal resistance, particle clustering and particle concentration, the fractal and chemical dimensions are fixed at 1.8 and 1.4 respectively. The values are in line with the work of Evans et al (2008).

**i.  Effect of interfacial thermal resistance ($\alpha$)**

The enhancement ratio ($k_{eff}/k_f$) was calculated for different cases of the average size of the nanoparticle cluster (in terms of $R_g/a$) for a range of interfacial thermal resistance ($\alpha$) values, i.e., $0.01 \leq \alpha \leq 10$, for two typical cases of volume fractions ($\emptyset_p$), 0.5% and 4%. These $\emptyset_p$ are commonly used in literature. Results are represented in Fig. 1.

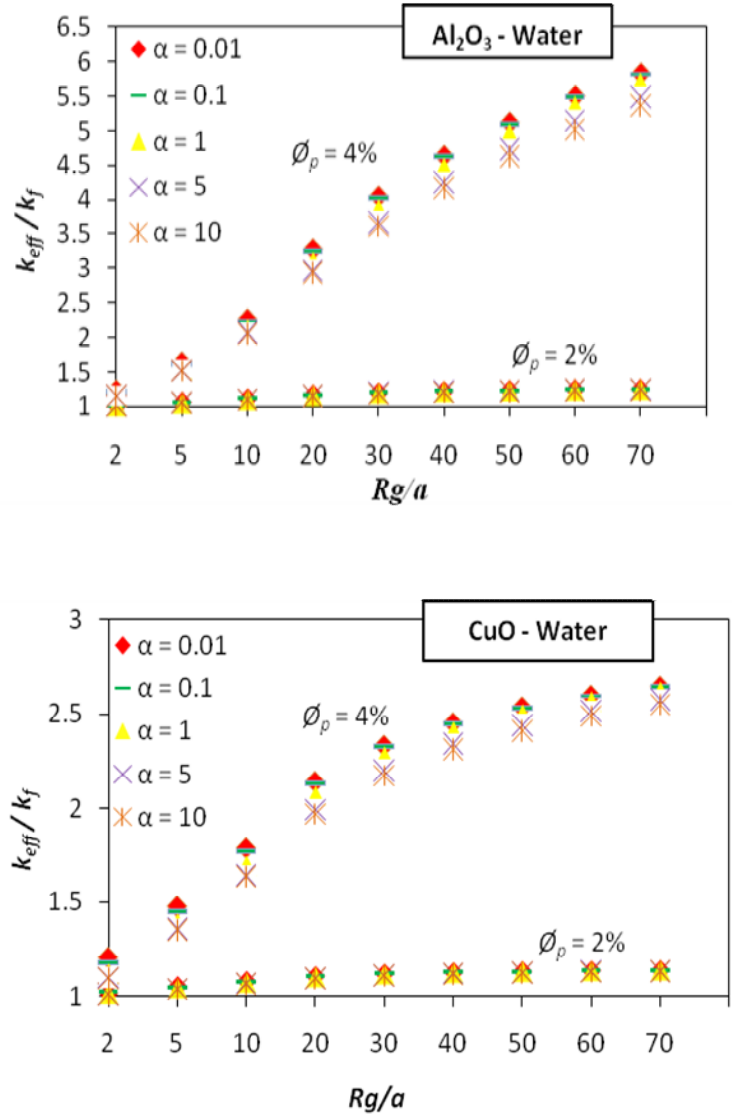

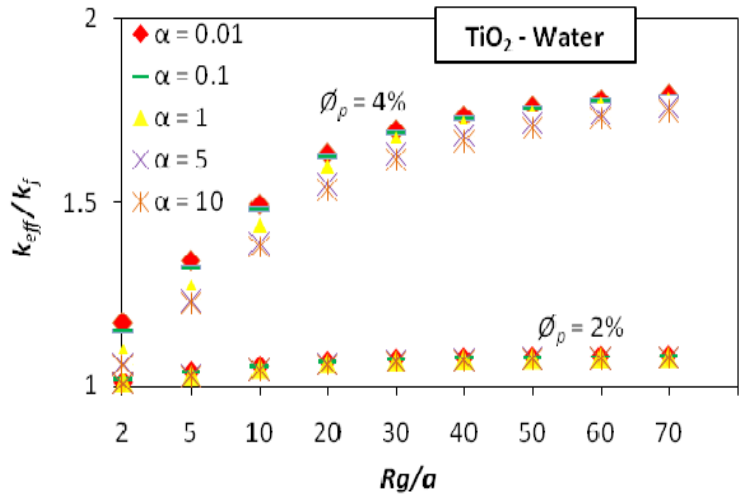

**Fig. 1** Dependence of $k_{eff}/k_f$ on $R_g/a$ and $α$. The bottom set of lines for 0.5vol%, the top set is for 4vol%

At 0.5vol% concentration, the $k_{eff}/k_f$ is fairly independent on $R_g/a$ as well as $α$. Moreover, as $φ_p$ increases to 4vol%, $α ≥ 1$ does not make a significant impact on enhancement ratio and so results of $α = 1$ (i.e. the radius of the nanoparticles is the same as the kapitza radius) is an average or median representation of this case of $α$ (i.e. $0.01 ≤ α ≤ 10$) being investigated. For the case of $α = 1$, the interfacial thermal resistance and therefore Kapitza radius plays no major role in thermal enhancement as its contribution is balanced by the higher thermal conductivity of the nanoparticles (Every *et al.*, 1992). These observations are consistent for all types of nanofluids shown in Fig. 1. Upon this observation, the value for $α$ was fixed to 1 in the following simulations.

ii. **Effect of aspect ratio ($Rg/a$) and particle volume fraction ($Ø_p$)**

Presented in Fig. 2 is the relationship between the enhancement ratio, aspect ratio and particle volume fraction. Note that $α$ is set to 1.

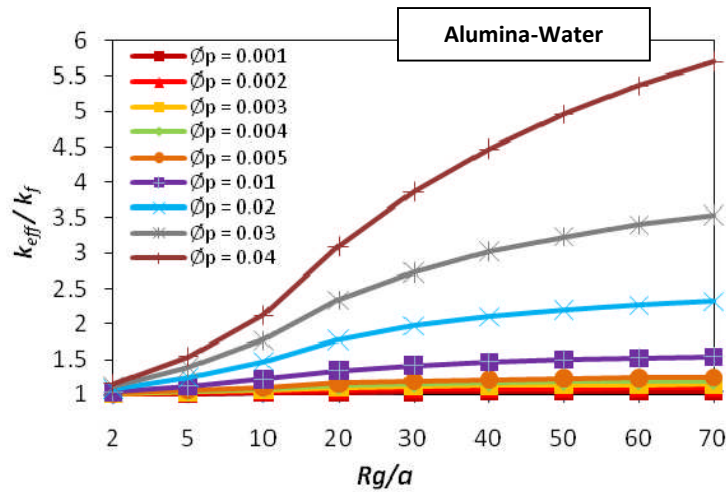

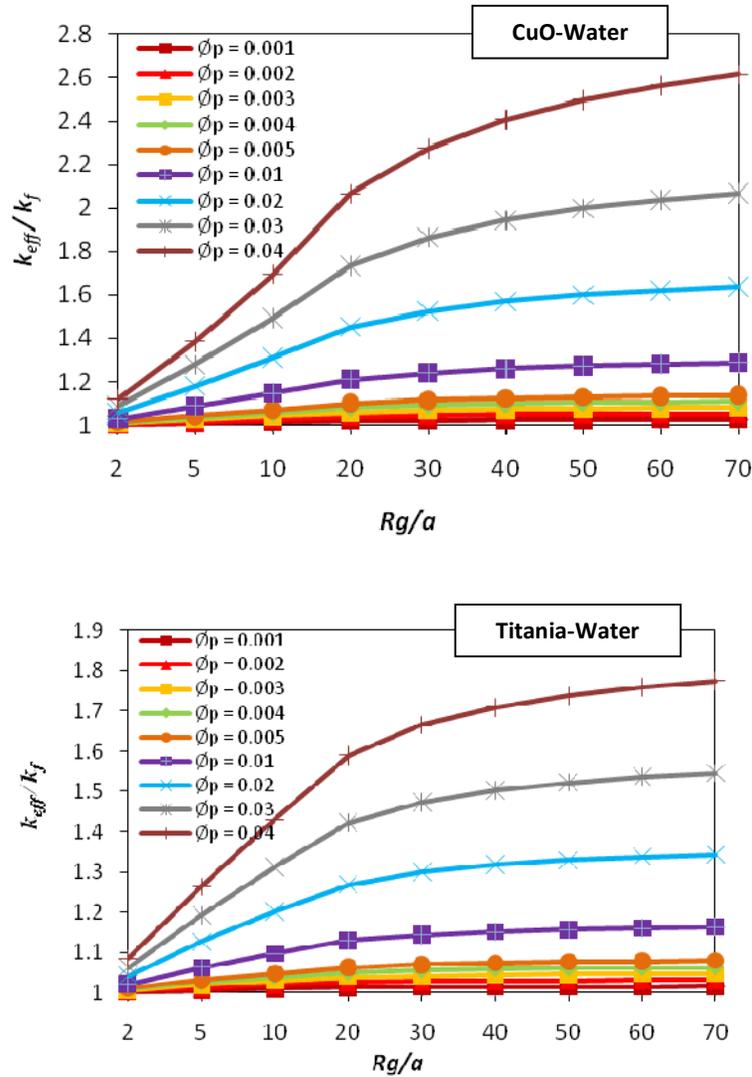

**Fig. 2** The effect of aspect ratio and particle volume fraction on enhancement ratio for α=1

The volume fractions chosen in Fig. 2 were chosen to have an upper limit of 4%. In most practical applications, this is thought to be the limiting cases, considering the clogging problem and pumping costs. It can be noted that a significant impact on enhancement ratio begins when the vol% >1. Moreover, there is a steep increase in enhancement ratio up to *Rg/a = 20*, after which the curves gradually flatten out. It is interesting to note that for stable nanofluids formulated from spherical primary particles, it has been known that *Rg/a* is often less than 10 (He et al., 2007; Wen and Ding, 2005).

### iii. Effect of fractal and chemical dimensions

The effect of fractal and chemical dimension is typically analyzed for water based nanofluids of Alumina for each value of aspect ratio between 2 and 8. For the reasons outlined before, $\alpha$ is set to 1.

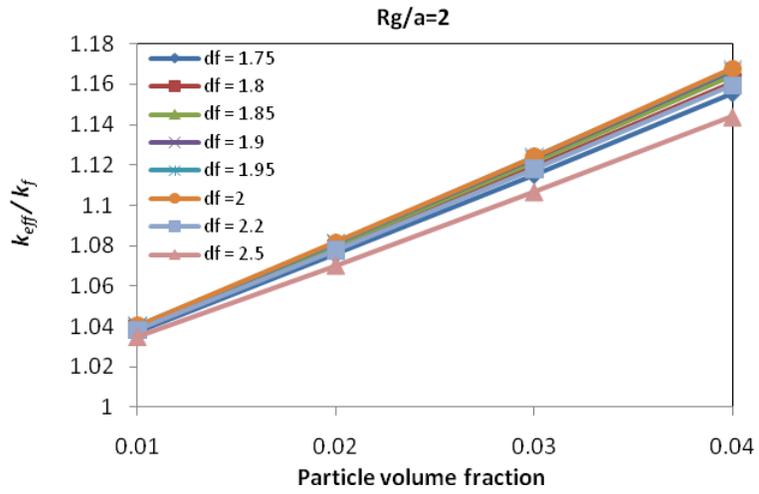

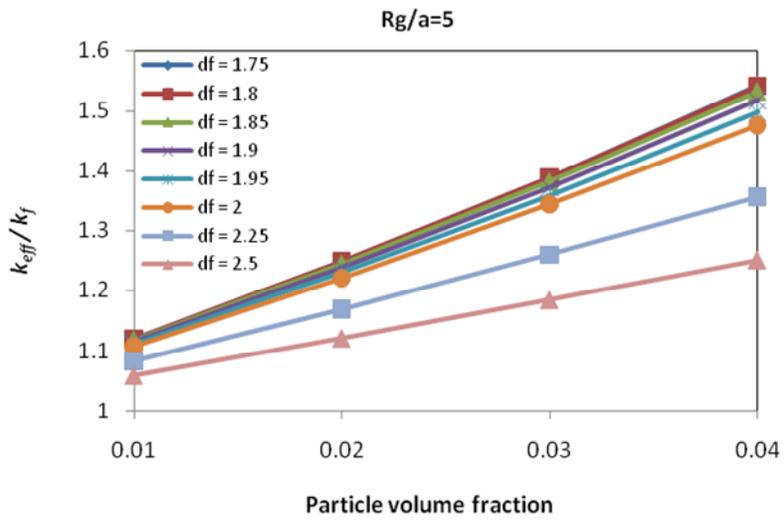

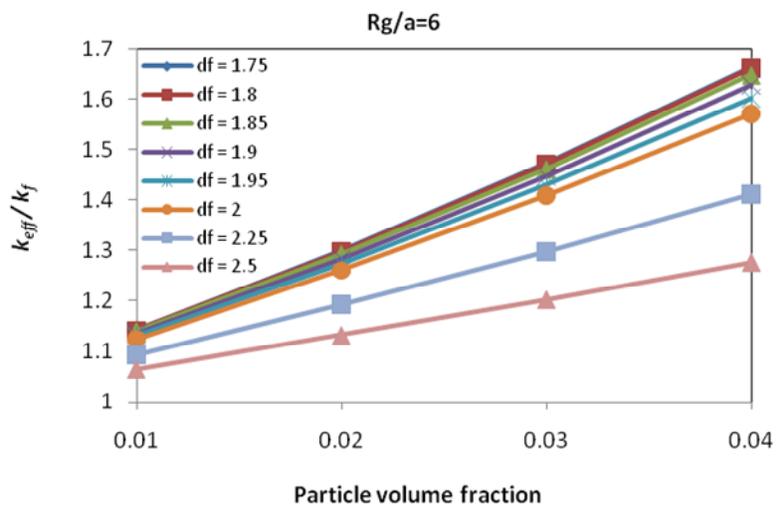

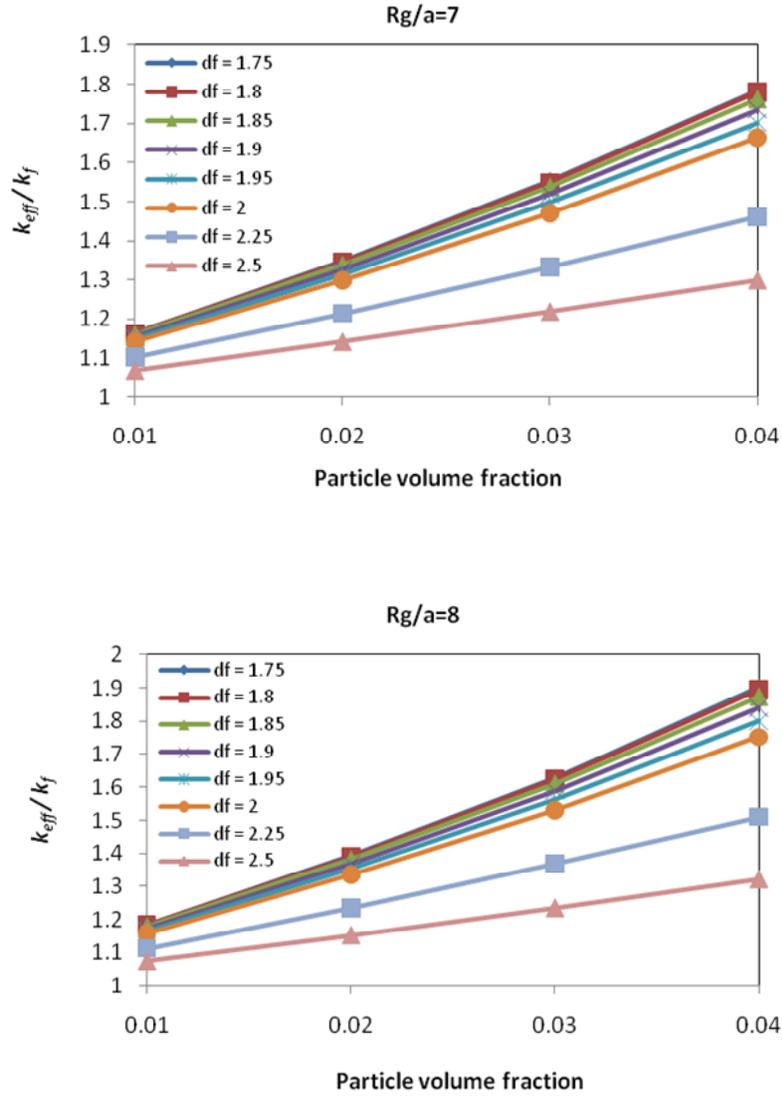

**Fig. 3** Effect of $d_f$ on $k_{eff}/k_f$ for Alumina-water nanofluid for $d_l = 1.4$ and $\alpha=1$

In the first instance from Fig. 3 above, the plot of enhancement ratio against particle volume fraction for different fractal dimension (i.e. ranging from about 1.8 to 2.5 and representing diffusion-limited cluster-cluster aggregation (DLCCA) for the lower limit and reaction-limited aggregation (RLA) for the upper limit of nanoparticles (Prasher *et al.*, 2006c) is shown. The plots reveal an increase in thermal enhancement for reducing fractal dimension as particle volume fraction increases. This observation also follows for increasing aspect ratio ($Rg/a$), in fact thermal enhancement increases. Wang et al (2003) has shown that nanofluid aggregation is DLCCA since the fractal dimensions lie close to 1.8. In the case of DLCCA, the aggregation process is known to be rapid thereby leading to loosely packed aggregates while for RLA, aggregation process is slower leading to the formation of densely packed aggregates (Waite *et al.*, 2001). A fitting of experimental data from literature in Fig. 3, will throw more light on the fractal dimension of nanofluids as well as the aggregation process (i.e. DLCCA or RLA).

Also considering the effect of chemical dimension, plots of enhancement ratio against particle volume fraction for different chemical dimension has been made for DLCCA systems ($d_f = 1.8$) as presented below in Fig. 4.

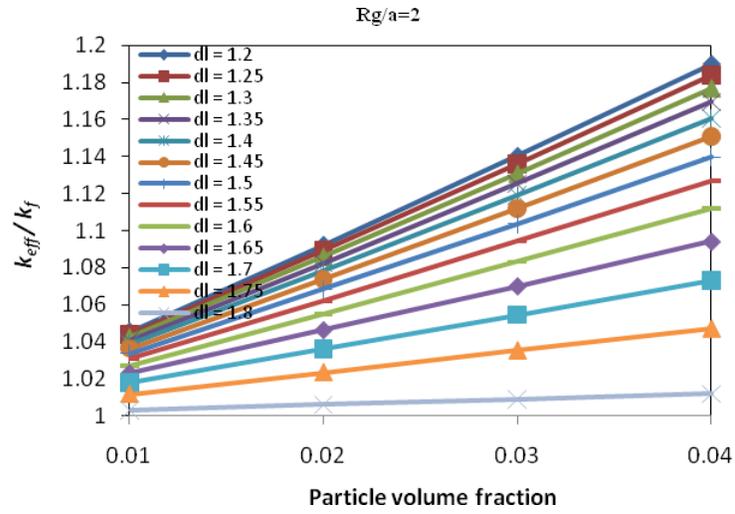

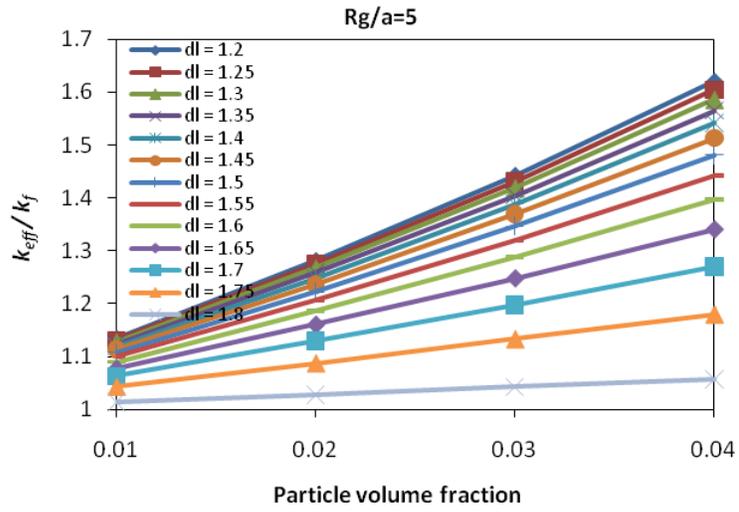

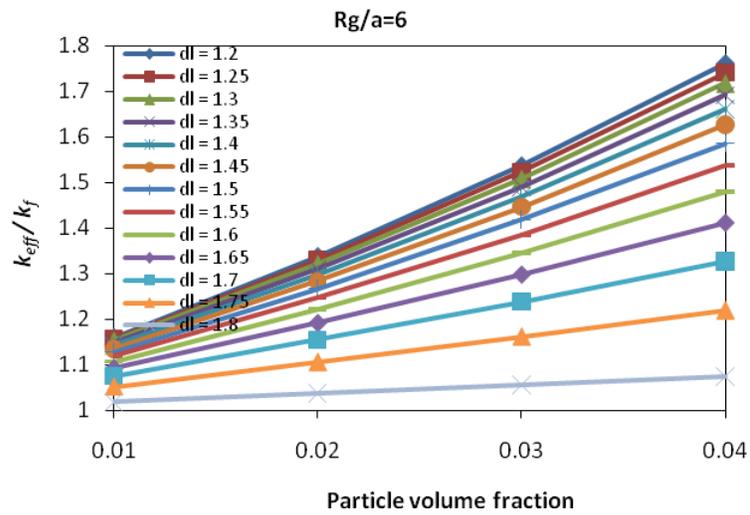

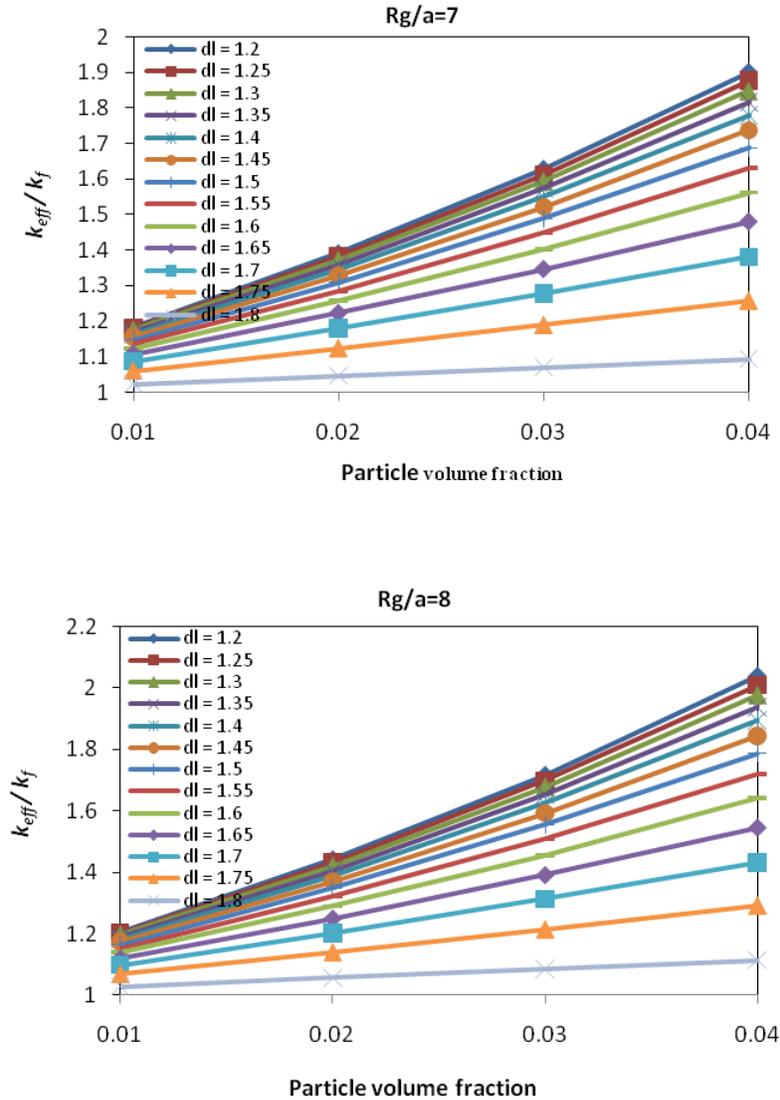

**Fig. 4** Effect of $d_l$ on $k_{eff}/k_f$ for Alumina-water nanofluid with for $d_f = 1.8$ and $\alpha = 1$

Following suit as it were in the case of fractal dimension, an increase in thermal enhancement for reducing chemical dimension and increasing particle volume fraction is observed. With a wide range of values of chemical dimension plotted, enhancement also increases as aspect ratio increases while chemical dimension reduces for increasing enhancement.

iv. **Validation of computational results with experimental data**

In the following plots we have validated the computational predictions of thermal enhancement of nanofluids with corresponding experimental values of selected nanofluid systems that are widely available in the literature. Initially we compared with the computational and experimental values of the aspect ratio ($Rg/a$) of nano-aggregates responsible for enabling enhancement of thermal conductance of the commonly used alumina-water nanofluid system as a model system. To validate our model, we collected experimental data on alumina-water nanofluids from the literature. There have been numerous publications on alumina-water nanofluids, for instance Avsec et al (2007), Feng et al (2007), Wang et al (2003), Yoo et al (2007), Murshed et al (2008), Ren et al (2005), Wong et al.(2008). However, only a few of them unambiguously specify primary particle size and the aggregate size. We have therefore recruited experimental datasets from Wen and Ding (2005) and Leong et al (2006).

Wen and Ding (2005) formulated Alumina-water nanofluids with particle loading up to 0.016 in volume fraction by dispersing Alumina nanopowder in water, followed by ultrasonication and homogenising. SEM images had confirmed that the primary particles were ranging in sizes of 10-50nm. After formulation, Malvern Nano ZS instrument was used to measure particle sizes. They found that the particle aggregates were having an average diameter of 167.5nm. Presumably they are aggregates of primary particles. To validate our model, values should be assigned to *Rg/a*. This needs estimated value range for primary particle size. This is chosen as in the range of 30-45nm, which agrees with experience. As a result, *Rg/a* will fall in an interval of 5.6 to 3.7 (Wen and Ding, 2005).

Leong et al (2006) had considered two particle sizes; 80nm and 150nm in their Alumina-water nanofluids of particle volume fractions up to 0.005. Formulation details were not detailed out but presumably by dispersing dry particles in water similar to their previous work (2005). Particle sizing methods were not clearly mentioned, but it is highly possible that they were referring to aggregate size. Here we recruit their experimental data for 150nm aggregate size. We assume that their primary particle size was in the interval of 30-45nm.

Fig. 5 demonstrates the model predictions and the experimental data as a function of $d_f$ for alumina-water nanofluid systems. For these cases, $d_l$ is taken as 1.4

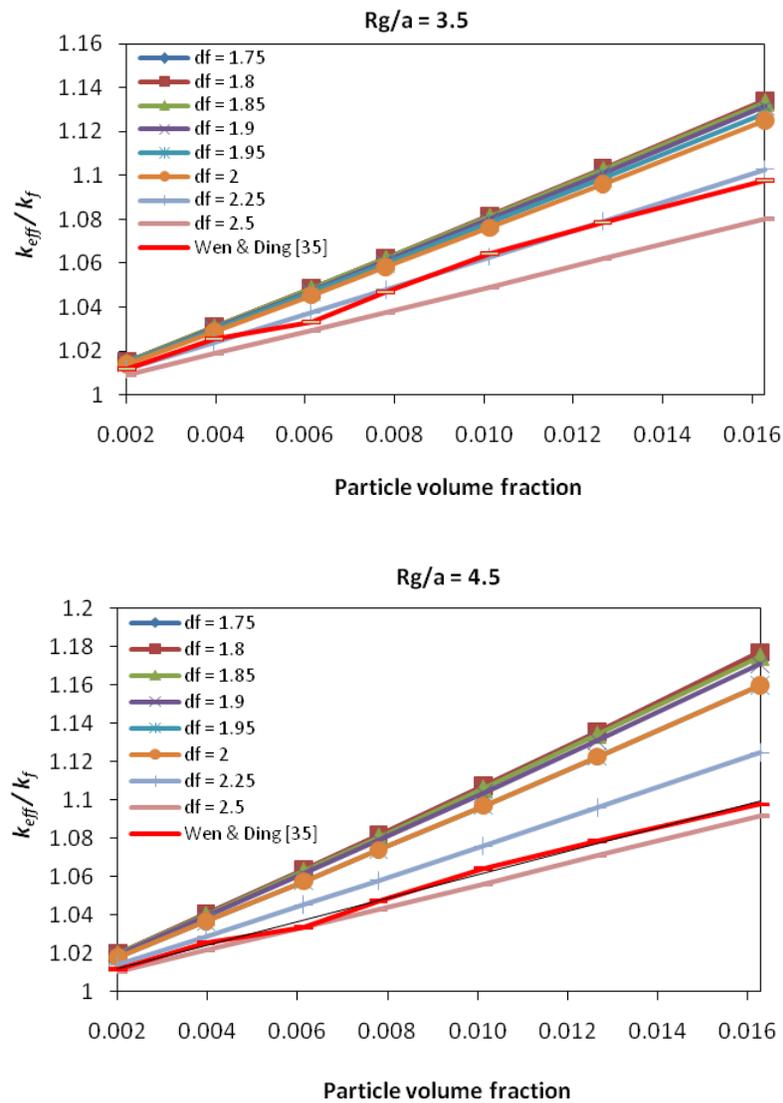

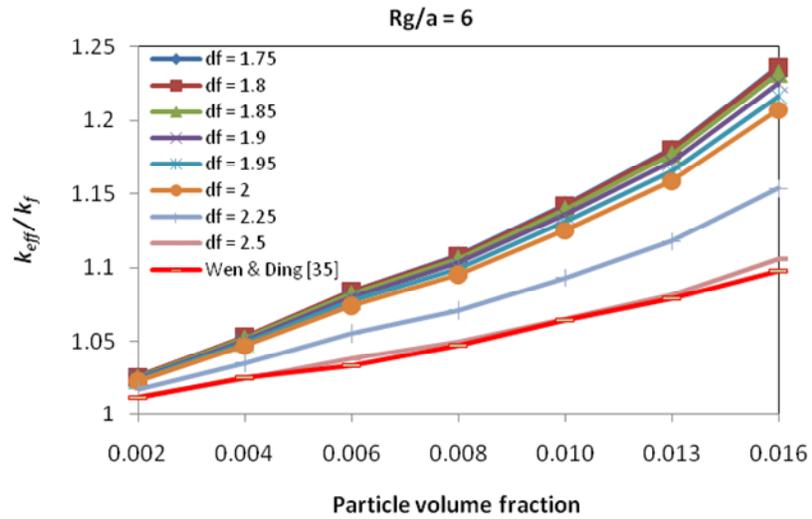

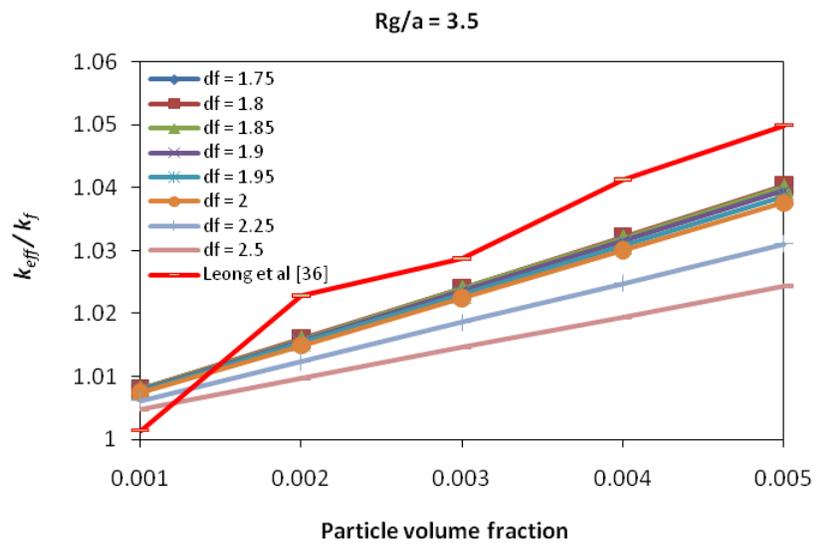

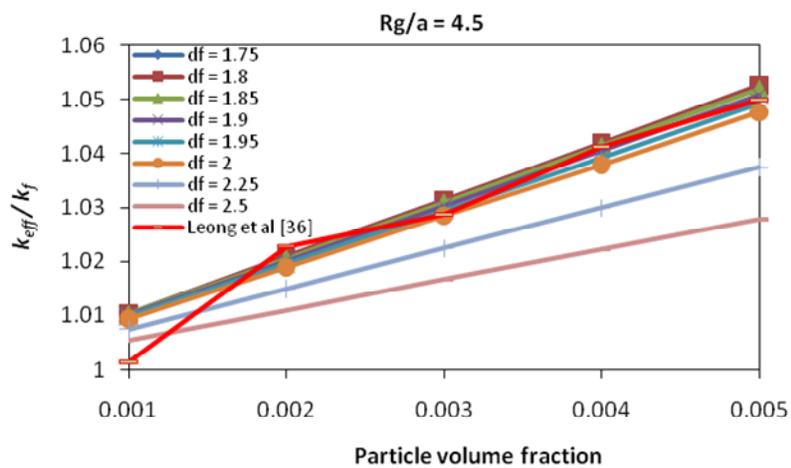

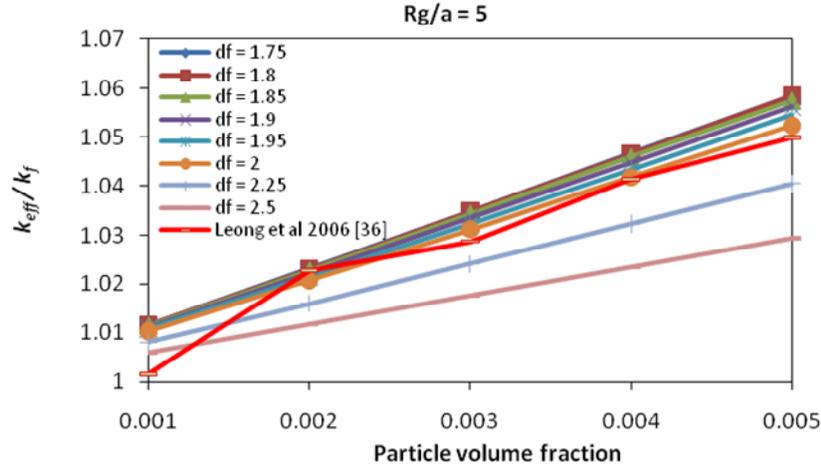

**Fig. 5** Comparison with literature. Alumina-water nanofluid with $d_l$=1.4 and α=1

It follows from Fig. 5 that the experimental data of Wen & Ding falls within the interval of 2.25< $d_f$ <2.5 while that of Leong et al falls within the interval of 1.75< $d_f$ <2. It can be speculated that the aggregation process in Wen & Ding is likely to be RLA induced while that of Leong et al, DLCCA. Recall that this was the dataset where primary particle size was clearly specified. This observation could be linked to a previous work of Waite et al (2001) where investigations were carried out on alumina-nanofluid system. Their work suggests that although the fractal dimensions obtained fell within the expected range of 1.8 to 2.3, it did not obey the conventional rule of DLCCA for lower fractal dimensions and RLA for higher fractal dimensions as they observed slightly higher fractal dimension for higher aggregation rates. This could be the reason for the observed closeness in thermal enhancement ratio (<1% difference) of experimental data. At the same time these data fall at different regions of fractal dimensions. Furthermore it is observed that the experimental data for Leong et al approaches the RLA interval as $Rg/a$ increases. This may suggest that the aggregation process gets slower with increasing aggregate size.

To further demonstrate the accuracy of present simulation technique, our data is compared with experimental results available in the literature for water based nanofluids of Alumina, CuO and Titania, shown in Fig. 6. The aspect ratio for nanoparticles ($Rg/a$) was fixed at a value of 2 assuming little or no aggregation (i.e. $Rg/a$ = 2) of particles for Alumina-water system. This is to compensate for the often undisclosed primary particle sizes in literature. For other systems, Rg/a varies between 5-8. Also following above analysis on fractal and chemical dimensions, $d_f$=1.8 and $d_l$=1.4.

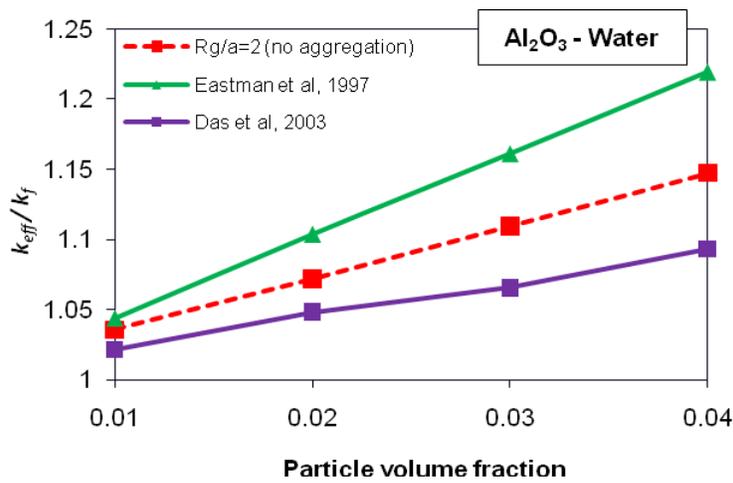

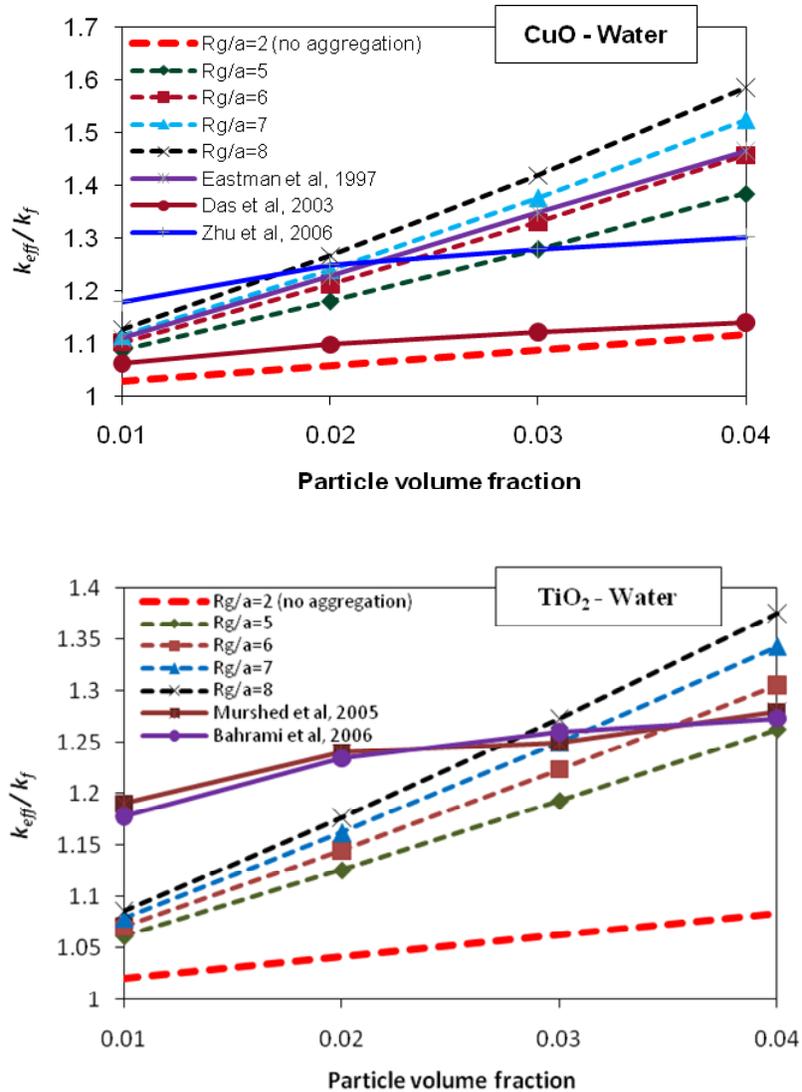

**Fig. 6** Comparison with literature data, $d_f = 1.8$, $d_l = 1.4$ and $α = 1$.

It can be seen that our models are in good agreement with experimental data from literature for aqueous Alumina and copper oxide nanofluids. For the Titania nanofluid, the model predictions follow the same trend as the experimental data fairly well.

**CONCLUSIONS**

The present work clearly suggests that aggregate size (reflected by $Rg/a$) has influence on thermal conductivity enhancement. This influence is significant when $Rg/a < 20$. The influence of interfacial thermal resistance, which has been accounted for by $α$ which reflects Kapitza radius was investigated, and it was observed that $α = 1$ is an average representation or median of this case of $α$ (i.e. $0.01 \leq α \leq 10$) being investigated. It is revealed that interfacial thermal resistance of $α = 1$ does not play a major role in thermal enhancement. Moreover, to achieve a sensible enhancement in thermal conductivity, particle concentration in the nanofluid is suggested to be above 1vol%.

Furthermore, the work also gives an insight into the effect of fractal and chemical dimension on thermal conductivity of nanofluids. Based on the aforementioned analysis and observations with water based alumina nanoparticles as the nanofluid, the wide range of values analyzed shows that both fractal and chemical dimensions reduce for increasing thermal enhancement. Also, validations carried out on fractal dimensions with experimental data from literature on alumina-water based nanofluids shows that there is an inter-link between aggregation mode (i.e. DLCCA or RLA), aggregate size and compactness of the aggregates, and thermal enhancement of the surrounding nanofluid. Aggregation mode influences the aggregation rate and compactness of the aggregates and also thermal enhancement. Validations show that the mode of aggregation of nanoparticles plays a major role in thermal enhancement irrespective of the type of nanofluid been investigated. As these observations occur for increasing particle volume fractions, the effect of fractal and chemical dimensions need not be underestimated when probing possible factors influencing thermal conductivity of nanofluids.

In summary, this study has shown how sensitive thermal conductivity is to aggregate sizes, particle concentrations, interfacial resistances including fractal and chemical dimensions. Still bearing in mind particle aggregation as a possible mechanism behind thermal enhancement, plots and results show that such parameters as fractal and chemical dimensions, work in conjunction with aggregation. These findings narrow down the search for possible mechanisms influencing thermal enhancement in nanofluids, and brings the end to this search closer. This will pave way for an in depth research into making adequate and proper use of these observed enhancement and efficiency in nanofluids, while harnessing them for the benefit of industries and other relevant sectors.

Further studies are being conducted on the influence of the type of interaction on self-assembly and mechanical strength of nanoparticles using theories of contact mechanics. It is aimed to predict more rigorously the interaction forces between nanoparticles in liquid environments and to accurately estimate the properties of clusters and their contribution to thermal enhancement of nanofluids.

## NOMENCLATURE

| | |
|---|---|
| $a_k$ | Kapitza radius |
| $a$ | radius of primary nanoparticles |
| $d_l$ | chemical dimension |
| $d_f$ | fractal dimension |
| $k$ | thermal conductivity |
| $N$ | number of particles |
| $R_g$ | radius of gyration |

*Greek symbols*
| | |
|---|---|
| $\varphi$ | volume fraction |

*Subscripts*
| | |
|---|---|
| $a$ | aggregate |
| $c$ | backbone |
| $eff$ | effective |
| $f$ | fluid |
| $nc$ | dead ends |
| $p$ | particle |

# REFERENCES

Avsec, J., and Oblak, M. (2007). The calculation of thermal conductivity, viscosity and thermodynamic properties for nanofluids on the basis of statistical nanomechanics. . International Journal of Heat and Mass Transfer *50*.

Bruggeman, D.A.G. (1935). Calculation of various physics constants in heterogenous substances I Dielectricity constants and conductivity of mixed bodies from isotropic substances. Annalen der Physik *24*, 636-664.

Chen, H., Yang, W., He, Y., Ding, Y., Zhang, L., Tan, C., Lapkin, A.A., and Bavykin, D.V. (2008). Heat transfer and flow behaviour of aqueous suspensions of titanate nanotubes (nanofluids). Powder Technology *183*, 63-72.

Ding, Y., Chen, H., Musina, Z., Jin, Y., Zhang, T., Witharana, S., and Yang, W. (2010). Relationship between the thermal conductivity and shear viscosity of nanofluids. Physica Scripta *T139*.

Evans, W., Prasher, R., Fish, J., Meakin, P., Phelan, P., and Keblinski, P. (2008). Effect of aggregation and interfacial thermal resistance on thermal conductivity of nanocomposites and colloidal nanofluids. International Journal of Heat and Mass Transfer *51*, 1431-1438.

Every, A.G., Tzou, Y., Hasselman, D.P.H., and Raj, R. (1992). The effect of particle size on the Thermal conductivity of ZnS/Diamond Composites. Acta metall mater *40*, 123-129.

Feng, Y., Yu, B., Xu, P., and Zou, M. (2007). The effective thermal conductivity of nanofluids based on the nanolayer and the aggregation of nanoparticles. Journal of Physics D: Applied Physics *40*.

Garg, P., Alvarado, J.L., Marsh, C., Carlson, T.A., Kessler, D.A., and Annamalai, K. (2009). An experimental study on the effect of ultrasonication on viscosity and heat transfer performance of multi-wall carbon nanotube-based aqueous nanofluids. International Journal of Heat and Mass Transfer *52*, 5090-5101.

Hamilton, R.L., and Crosser, O.K. (1962). Thermal conductivity of hetrogeneous two-component systems. Industrial & Engineering chemistry fundamentals *1*, 187-191.

He, Y.R., Jin, Y., Chen, H.S., Ding, Y.L., Cang, D.Q., and Lu, H.L. (2007). Heat transfer and flow behaviour of aqueous suspensions of TiO2 nanoparticles (nanofluids) flowing upward through a vertical pipe. International Journal of Heat and Mass Transfer *50*, 2272-2281.

J. Buongiorno, D.C. Venerus, N. Prabhat, T. McKrell, J. Townsend, R. Christianson, Y.V. Tolmachev, P. Keblinski, L.W. Hu, J. L. Alvarado*, et al.* (2009). A benchmark study on the thermal conductivity of nanofluids. Journal of Applied Physics *106*.

Jang, S.P., and Choi, S.U.S. (2004). Role of Brownian motion in the enhanced thermal conductivity of nanofluids. Applied Physics Letters *84*, 4316-4318.

Keblinski, P., Phillpot, S.R., Choi, S.U.S., and Eastman, J.A. (2002). Mechanisms of heat flow in suspensions of nano-sized particles (nanofluids). International Journal of Heat and Mass Transfer *45*, 855-863.

Kim, S.J., Bang, I.C., Buongiorno, J., and Hu, L.W. (2007). Surface wettability change during pool boiling of nanofluids and its effect on critical heat flux. International Journal of Heat and Mass Transfer *50*, 4105-4116.

Koo, J., and Kleinstreuer, C. (2004). A new thermal conductivity model for nanofluids. Journal of Nanoparticle Research *6*, 577-588.

Kumar, D.H., Patel, H.E., Kumar, V.R.R., Sundararajan, T., Pradeep, T., and Das, S.K. (2004). Model for heat conduction in nanofluids. Physical Review Letters *93*.

Leong, K.C., Yang, C., and Murshed, S.M.S. (2006). A model for the thermal conductivity of nanofluids - the effect of interfacial layer. Journal of Nanoparticle Research *8*, 245-254.

Li, C.H., and Peterson, G.P. (2006). Experimental investigation of temperature and volume fraction variations on the effective thermal conductivity of nanoparticle suspensions (nanofluids). Journal of Applied Physics *99*.

Margolina, A. (1985). The fractal dimension of cluster perimeters generated by a kinetic walk. . Journal of Physics A: Mathematical and General *Vol. 18*.

Maxwell, J.C. (1881). A treatise on Electricity and Magnetism, Vol 1, 2 edn (Clarendon Press, Oxford, UK).

Murshed, S.M.S., Leong, K.C., and Yang, C. (2005). Enhanced thermal conductivity of TiO2 - water based nanofluids. International Journal of Thermal Sciences *44*, 367-373.

Murshed, S.M.S., Leong, K.C., and Yang, C. (2008). Thermophysical and electrokinetic properties of nanofluids – A critical review. Applied Thermal Engineering *28*.

Nan, C.W., Birringer, R., Clarke, D.R., and Gleiter, H. (1997). Effective thermal conductivity of Particulate composites with inetrfacial themal resistance. Journal of Applied Physics *81*, 6692-6699.

Prasher, R., Bhattacharya, P., and Phelan, P.E. (2005). Thermal conductivity of nanoscale colloidal solutions (nanofluids). Physical Review Letters *94*.

Prasher, R., Bhattacharya, P., and Phelan, P.E. (2006a). Brownian-motion-based convective-conductive model for the effective thermal conductivity of nanofluids. Journal of Heat Transfer-Transactions of the Asme *128*, 588-595.

Prasher, R., Evans, W., Meakin, P., Fish, J., Phelan, P., and Keblinski, P. (2006b). Effect of aggregation on thermal conduction in colloidal nanofluids. Applied Physics Letters *89*.

Prasher, R., Phelan, P.E., and Bhattacharya, P. (2006c). Effect of aggregation kinetics on the thermal conductivity of Nanoscale colloidal solutions (Nanofluids). Nano Letters *6*, 1529-1534.

Putnam, S.A., Cahill, D.G., Braun, P.V., Ge, Z.B., and Shimmin, R.G. (2006). Thermal conductivity of nanoparticle suspensions. Journal of Applied Physics *99*.

Ren, Y., Xie, H., and Cai, A. (2005). Effective thermal conductivity of nanofluids containing spherical nanoparticles. Journal of Physics D-Applied Physics *38*, 3958-3961.

Waite, T.D., Cleaver, J.K., and Beattie, J.K. (2001). Aggregation Kinetics and Fractal Structure of gamma-Alumina Assemblages. Journal of Colloid and Interface science *241*, 333-339.

Wang, B.-X., Zhou, L.-P., and Peng, X.-F. (2003). A fractal model for predicting the effective thermal conductivity of liquid with suspension of nanoparticles. International Journal of Heat and Mass Transfer *46*.

Wen, D.S., and Ding, Y.L. (2004). Experimental investigation into convective heat transfer of nanofluids at the entrance region under laminar flow conditions. International Journal of Heat and Mass Transfer *47*, 5181-5188.

Wen, D.S., and Ding, Y.L. (2005). Experimental investigation into the pool boiling heat transfer of aqueous based gamma-alumina nanofluids. Journal of Nanoparticle Research *7*, 265-274.

Wen, D.S., Ding, Y.L., and Williams, R.A. (2006). Pool boiling heat transfer of aqueous TiO2-based nanofluids. Journal of Enhanced Heat Transfer *13*, 231-244.

Witharana, S., Chen, H., and Ding, Y. (2010). Shear stability and thermophysical properties of nanofluids. In IEEE-International conference on Information and Automation for Sustainability (Colombo, Sri Lanka).

Wong, K.V., and Kurma, T. (2008). Transport properties of alumina nanofluids. Nanotechnology *19*.

Xuan, Y., and Li, Q. (2000). Heat transfer enhancement of nanofluids. International Journal of Heat and Fluid flow *21*, 58-64.

Xuan, Y., Li, Q., and Hu, W. (2003). Aggregation structure and thermal conductivity of nanofluids. AIChE Journal *49*, 1038-1043.

Yoo, D., Hong, K.S., and Yang, H. (2007). Study of thermal conductivity of nanofluids for the application of heat transfer fluids. Thermochimica Acta *455*.